\documentstyle[pre,aps,]{revtex}

\newcommand{\beq}{\begin{equation}}
\newcommand{\eeq}{\end{equation}}
\newcommand{\bqn}{\begin{eqnarray}}
\newcommand{\eqn}{\end{eqnarray}}
\newcommand{\bqns}{\begin{eqnarray*}}
\newcommand{\eqns}{\end{eqnarray*}}
\newcommand{\bary}{\begin{array}}
\newcommand{\eary}{\end{array}}
\newcommand{\non}{\nonumber}
\input{epsf}

\begin{document}
\twocolumn[\hsize\textwidth\columnwidth\hsize\csname %
@twocolumnfalse\endcsname
\title{Poincar\'{e} cycle of a multibox Ehrenfest urn model with
directed transport}
\author{Yee-Mou Kao}
\address{National Center for High-Performance Computing,
No.21, Nan-ke 3rd.Rd., Hsin-Shi, Tainan County 744, Taiwan,
Republic of China}
\author{Pi-Gang Luan}
\address{Institute of Optical Sciences, National Central
University, Chung-Li 32054, Taiwan, Republic of China}
\date{\today}
\maketitle
\begin{abstract}
We propose a generalized Ehrenfest urn model of many urns arranged
periodically along a circle. The evolution of the urn model system
is governed by a directed stochastic operation. Method for solving
an $N$-ball, $M$-urn problem of this model is presented. The
evolution of the system is studied in detail. We find that the
average number of balls in a certain urn oscillates several times
before it reaches a stationary value. This behavior seems to be a
peculiar feature of this directed urn model. We also calculate the
Poincar\'{e} cycle, i.e., the average time interval required for
the system to return to its initial configuration. The result can
be easily understood by counting the total number of all possible
microstates of the system.
\vspace{1cm}
\end{abstract}
]

\section{introduction}
Physical laws governing the microscopic processes are mostly
reversible in time. In macroscopic world, however, people often
experience time-irreversible phenomena in their daily life. To
understand why the reversible microscopic processes lead to
irreversible macroscopic manifestations one refers to the {\it
Poincar\'{e} Theorem}, which states that a system having a finite
energy and confined to a finite volume will, after a sufficient
long time -- the so called {\it Poincar\'{e} cycle}, return to an
arbitrarily small neighborhood of almost any given initial state
\cite{Huang}. The key point is to note that the typical value of a
Poincar\'{e} cycle for even a moderate-sized system is far beyond
the meaningful time scale one can measure or experience, thus the
irreversiblity is realized.

Usually to describe a macroscopic system one has to know only a
few parameters, such as volume, pressure, and temperature.
However, to describe the same system in terms of its microscopic
constituents, one has to deal with a large number of parameters,
such as the momenta and positions of a huge amount of particles,
which are impossible to calculate in practice. Based on this
reason together with the fact that the macroscopic laws are
insensitive to the microscopic details (of system history), it is
natural for people to adopt the probability (ensemble) description
in statistical mechanics, which deals with the equilibrium state
(a macroscopic state that has stationary value of state
parameters) of a macroscopic system. In this kind of description
the macroscopic quantities are defined as the ensemble average of
their microscopic correspondences. This definition connects the
microscopic and macroscopic worlds.

To study how a system approaches its equilibrium state one also
uses probability description, where the evolution of the system is
treated as a stochastic process. One famous model for simulating
such a process was proposed by Ehrenfest one century ago
\cite{Erf}, which is an $N$-ball, $2$-urn problem. In the
beginning $N$ numbered balls are distributed arbitrarily in either
urn $A$ or urn $B$. At each time step one ball is picked out at
random and then put into the other urn. This simple model can be
exactly solved to give an explicit Poincar\'{e} cycle. This model
was then generalized by several authors to mimic more complicated
situations encountered in real physical phenomena
\cite{Siegert,Klein,Calvo}. An attractive feature of these urn
model problems is that they are easy to formulate but not always
easy to solve. The solutions obtained have, therefore, sometimes
led to new mathematical techniques and insights
\cite{God,Kanter,Lipo,Arora}. Recently, some new urn models were
proposed and solved analytically or numerically. Their results
provide very good descriptions on granular and glass systems
\cite{Ritort,Eggers,Na,van,Droz}.

\begin{figure}[hbt]
\epsfxsize=2.8in\epsffile{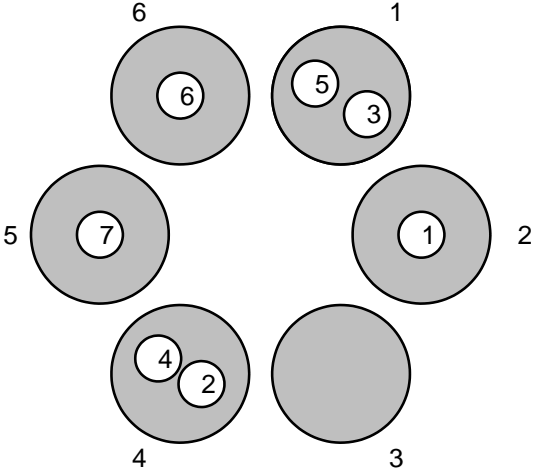} \caption{Arrangement of
the numbered urns and balls in our periodic urn model. The gray
disks represent the urns, and the white disks represent the balls.
Here we illustrate a configuration for a system with six urns and
seven balls. The state vector for this configuration is $|{\bf
m}\rangle=|2,1,0,2,1,1\rangle$.}\label{arg}
\end{figure}

In this paper, we obtain the exact solution of a generalized urn
model. Hereafter we call it ``periodic urn model". In this model,
one considers $N$ distinguishable balls which are distributed in
$M$ urns. These $M$ urns are arranged along a circle and numbered
one by one to form a cycle, that is, we define the $(M+1)$th urn
as the 1st urn (See Fig.~\ref{arg}). To begin with, the initial
distribution of the $N$ balls in the $M$ urns is given by $
|m_{1,0},\,m_{2,0},\,\cdots,\,m_{M,0} \rangle \equiv |{\bf  m}_0
\rangle $, where $m_{i,0}$ is the number of balls in the $i$th urn
at the start. At each time step one ball is picked out of the $N$
balls such that every ball has an equal probability of being
picked up. The ball is then placed into the next numbered urn. The
state that the $i$th urn contains $m_i$ balls is represented by $|
m_1,m_2,\cdots,m_M \rangle  \equiv | {\bf  m} \rangle $, which we
name it {\it state vector}. Hereafter we call a distribution
string ${\bf m}$ (without knowing the numbering of the balls) a
{\it configuration} of the system. Otherwise, if we also know the
location of each numbered ball, we call such a distribution a {\it
microstate} of the system.

After $s$ steps, the transition probability from state $|{\bf
m}_0\rangle$ to state $|{\bf m}\rangle$ can be written as $\langle
{\bf m}|P^s|{\bf m}_0\rangle$, where $P$ represents the operation
in one step, and the orthonomality conditions of these state
vectors have been assumed.

According to the above description, the transition probabilities
corresponding to the $s$th step and the $(s-1)$th step satisfy the
recursion relation \bqn
   &&\langle m_1,m_2,\cdots,m_M|P^s|{\bf  m}_0\rangle \non\\
    &&=\sum_{i=1}^{M}\frac{m_i+1}{N}\langle \cdots,m_i+1,m_{i+1}-1,
    \cdots|P^{s-1}|{\bf  m}_0\rangle,\label{recu}
\eqn where $m_{M+1}=m_1$ as has been mentioned before. In
addition, any state that does not satisfy the constraint \beq
m_1+m_2+\cdots+m_M=N.\label{constraint} \eeq is an unphysical
state and has null contribution in the sum. Hereafter we will use
Eq.~(\ref{recu}) as the basis to derive all the results we want to
know.

This paper is organized as follows.  In the next section we
calculate the average number of balls in an urn at any time. In
section III we introduce a generating function that has $N$
variables and solve the problem completely. In section IV the
solution of the model will be applied to the calculation of
Poincar\'{e} cycle. Finally in section V we give the summary of
this paper.

\section{Average number of balls in an urn}

The first thing we want to know is how many balls on average will
appear in the $i$th urn after first $s$ steps. We define the
average (the expectation value) of a quantity $A$ (which depends
on the state vector $|{\bf m}\rangle$ at each step, written as
$A({\bf m})$) after $s$ steps as \beq
 \langle A  \rangle _s = \sum_{\{{\bf  m}\}} A({{\bf  m}})\langle
 {\bf  m}|P^s|{\bf  m}_0\rangle ,\label{expect}
\eeq where $\{{\bf  m}\}$ include all the configurations
satisfying the constraint (\ref{constraint}).

Let $A({{\bf m}})=m_i$, then from Eq.~(\ref{recu}) and
(\ref{expect}) we have \bqn \langle m_i\rangle _s &=& \sum_{\{{\bf
m}\}} m_i\langle {\bf m}|P^s|{\bf m}_0\rangle , \non\\
 &=& \sum_{\{{\bf  m}\}}\sum_{j=1}^{M}\frac{m_i (m_j+1 )}{N}\non\\
  & &\times  \langle \cdots,m_j+1,m_{j+1}-1,\cdots|P^{s-1}|{\bf  m}_0\rangle,\non \\
 &=& \sum_{j=1}^{M}
       \left( \frac{\langle m_i m_j\rangle _{s-1}}{N}
             -\frac{\langle m_j\rangle _{s-1}}{N}\delta_{j,i}
             +\frac{\langle m_j\rangle _{s-1}}{N}\delta_{j+1,i} \right),\non\\
 &=& \sum_{j=1}^{M}\frac{\langle m_i m_j\rangle _{s-1}}{N}
 -\frac{\langle m_i\rangle _{s-1}}{N}+\frac{\langle m_{i-1}\rangle _{s-1}}{N},\non\\
&=& \left(1- \frac{1}{N}\right)\langle m_i\rangle
_{s-1}+\frac{\langle m_{i-1}\rangle _{s-1}}{N}, \label{mm}\eqn
here we have used the constraint (\ref{constraint}).

Now we are ready to solve $\langle m_i \rangle_s$. The recurrence
relation (\ref{mm}) can be written as \beq {\cal M}_s= P_{ave}
{\cal M}_{s-1}, \label{recmm} \eeq where ${\cal M}_s$ is a $M
\times 1$ column vector defined by
 \beq
 {\cal
 M}_s= \left[ \bary{cccc} \langle m_1\rangle _s  \\
 \langle m_2\rangle _s  \\
 \vdots  \\ \langle
 m_M\rangle _s \eary \right],\eeq and $P_{ave}$ is a $ M \times M
 $ matrix written as \beq
 P_{ave}= \left[ \bary{cccc} 1- \frac{1}{N} & 0 & \cdots & \frac{1}{N} \\
   \frac{1}{N} & 1- \frac{1}{N} &  \cdots & 0  \\
                 \vdots &   \vdots      &   \ddots & \vdots      \\
         0     &    0          &  \cdots & 1- \frac{1}{N} \eary \right].
\eeq

\begin{figure}[hbt]
\epsfxsize=2.8in\epsffile{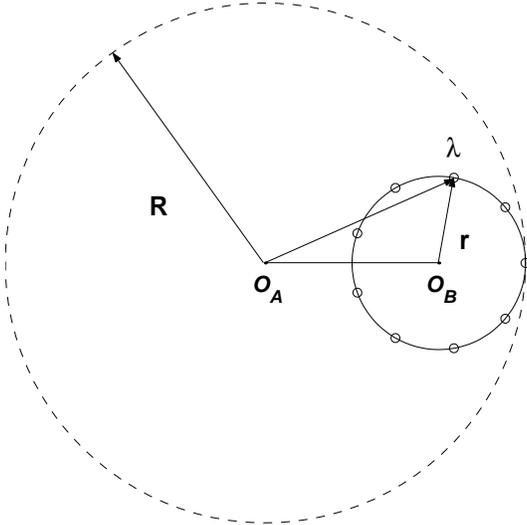} \caption{Eigenvalues
$\{\lambda_m\}$ of the matrix $P_{ave}$ (represented by the tiny
circles). Here $R=1$ and $r=1/N$ are the radii of two reference
circles, and $O_A$ and $O_B$ are their centers, respectively. The
eigenvalues of $P_{ave}$ are distributed uniformly on the small
reference circle centered at $O_B=(1-1/N,0)$. } \label{roots}
\end{figure}

By means of the recurrence relation (\ref{recmm}), ${\cal M}_s$
can be deduced: \beq
 {\cal M}_s =P_{ave}^s {\cal M}_0,\label{mpsm}
\eeq
where
\beq
   {\cal M}_0= \left[ \bary{cccc} m_{1,0} \\  m_{2,0} \\
    \vdots  \\  m_{M,0}  \eary \right]
\eeq represents the initial state. $P^s_{ave}$ can be calculated
if one knows the eigenvalues $\lambda_m$ and eigenvectors $Q_m$ of
$P_{ave}$. They are given by  (See Fig.~\ref{roots})
\beq
  \lambda_m = 1-\frac{1}{N}+\frac{1}{N}q_m^{*},\;\;\;\;\;
 Q_m = \frac{1}{\sqrt{M}}
 \left[ \bary{cccc}
  q_m  \\ q_m^2 \\ \vdots  \\  q_m^M
 \eary \right],\label{lambdaq}
\eeq where \beq
 q_m = \exp \left(\frac{2m\pi i}{M}\right),\;\;\;\;\;
 m=1,2,\cdots,M.\label{qm}
\eeq

Denote $R$ as the $M\times M$ matrix of the eigenvectors $Q_m$
\bqn
 R &=&[Q_1,Q_2,\cdots,Q_M]\non\\
  &&\non\\
 &=& \frac{1}{\sqrt{M}} \left[
 \bary{cccc}
 q_1    &  q_2       & \cdots & q_M   \\
 q_1^2  &  q_2^2    & \cdots & q_M^2 \\
 \vdots &  \vdots  & \ddots & \vdots
 \\ q_1^M  &  q_2^M  & \cdots & q_M^M \eary
 \right]
\eqn and $\Lambda$ as the diagonal matrix of $P_{ave}$'s
eigenvalues $\lambda_m$ \beq
 \Lambda = \left[
 \bary{cccc} \lambda_1 &  0 &  \cdots & 0   \\
 0   &
 \lambda_2 &  \cdots & 0 \\
 \vdots    & \vdots   & \ddots & \vdots \\
  0         &  0       & \cdots & \lambda_M \eary \right]
\eeq
then we obtain
\beq
 P^{s}_{ave}=R\Lambda^{s} R^{-1}=R\Lambda^{s} R^{\dag}=R\Lambda^{s} R^{*},
\eeq where we have used the following properties of $R$
\beq
 R=R^{t},\;\;(R^{-1})=R^{\dag}=(R)^{*},\;\;R_{mn}=q^m_n=q^{mn}_1 .
\eeq

Now the average number of balls in the $i$th urn after $s$ steps
can be determined: \bqn
 \langle m_i\rangle _s &=&
 \frac{1}{M}\sum_{j=1}^{M}\sum_{k=1}^{M}\sum_{l=1}^{M}
 q_j^i\lambda_j^s\delta_{jk}q_l^{-k}\langle m_l\rangle _0 ,
 \non \\
            &=& \frac{1}{M}\sum_{j=1}^{M} \sum_{l=1}^{M}
                q_1^{j(i-l)}\lambda_j^s
                m_{l,0},\label{recu1}
\eqn where $\langle m_l\rangle _0 = m_{l,0}$ is the initial number
of balls in the $l$th urn.

\begin{figure}[hbt]
\epsfxsize=3.2in\epsffile{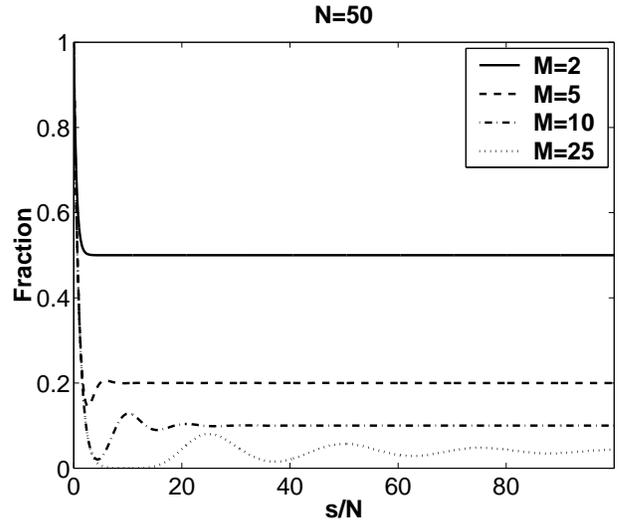} \caption{Average
number of balls in the first urn $\langle m_1\rangle_s$ as a
function of time $s$, assuming initially all the balls are in the
first urn. Here we plot ``Fraction"$=\langle m_1\rangle_s/N$ for
$N=50$ at $M=2,5,10,25$. As one can see, except for the $M=2$ case
(which is the original Ehrenfest urn model), this mean value
oscillates several times before it reaches a stationary value.}
\label{mean}
\end{figure}

Let us now consider a simple example. Suppose initially all the
$N$ balls are in the first urn, that is, \beq
 m_{1,0}=N,\;\;\; m_{2,0}=m_{3,0}= \cdots =
 m_{M,0}=0,
\eeq then according to Eq.~(\ref{recu1}), we have \beq
    \langle m_1\rangle _s = \frac{N}{M}
    \sum_{j=1}^{M}\lambda^s_j\label{mmean}
\eeq

Fig.~\ref{mean} shows the results for $N=50$ at $M=2,5,10,25$. The
$M=2$ case is the original Ehrenfest model, in which the average
number of balls in the first urn decays to $N/2$ in a period of
steps of order $N$. For any $M>2$ case, however, we observe that
before the system arrives its true equilibrium (here we mean the
value of $\langle m_i\rangle_s$ for each $i$ does not change
anymore), $\langle m_1\rangle _s$ undergoes several oscillations,
which seems to be a unique feature of this model and have never
been found in other kinds of urn models --- to our knowledge.
Furthermore, in Fig.~\ref{mean} the $M=25$ case shows that before
the appearance of the first peak of $\langle m_1\rangle _s$ there
is a period during which $\langle m_1\rangle _s$ is almost zero.
This phenomenon together with the oscillations mentioned before
seem to be typical results when both $M$ and $N$ are large.

\begin{figure}[hbt]
\epsfxsize=3.2in\epsffile{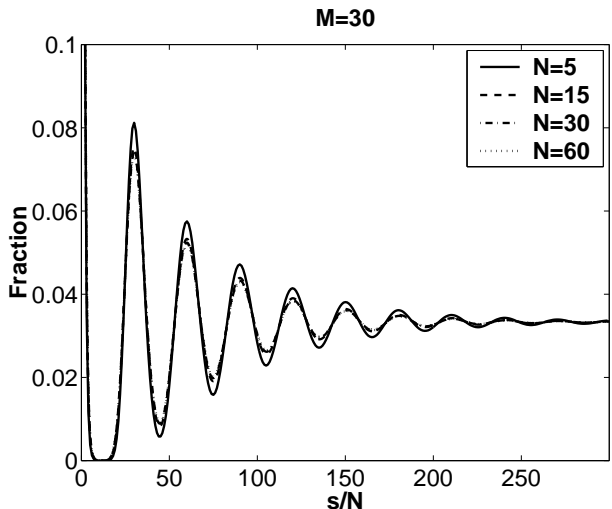} \caption{Plot of
``Fraction"$=\langle m_1\rangle_s/N$ as a function of time $s/N$,
assuming initially all the balls are in the first urn. Here $M=30$
and $N=5,15,30,60$. Except for the $N=5$ case (this $N$ is too
small), all the curves merge and become one. The visible range of
``Fraction " has been tuned to give a better illustration.}
\label{mean1}
\end{figure}

\begin{figure}[hbt]
\epsfxsize=3.2in\epsffile{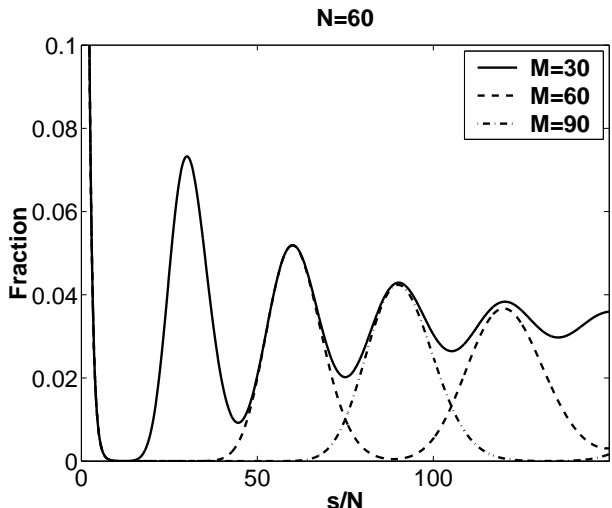} \caption{Plot of
``Fraction"$=\langle m_1\rangle_s/N$ as a function of time $s/N$,
assuming initially all the balls are in the first urn. Here $N=60$
and $M=30,60,90$. The first peaks of these three curves are
located at $s/N$=$30$, $60$, $90$, respectively.} \label{mean2}
\end{figure}

To understand the behaviors of $\langle m_1\rangle_s$ when $N$ is
large, note first that in this limit Eq.~(\ref{mmean}) can be
approximated by \beq
 \langle
 m_1\rangle _s  =  \frac{N}{M} \sum_{j=1}^{M}
       \exp\left[ \frac{s(q_j^{-1}-1)}{N}\right],\label{exp}
\eeq hence $\langle m_1\rangle_s/N$ becomes a universal function
of $s/N$ for a fixed $M$. In Fig.~\ref{mean1} we plot the $\langle
m_1\rangle_s/N$ curves for $M=30$ (which is large enough in
practice) and $N=5,15,30,60$. The $N=5$ case is shown here too to
compare with those cases with large $N$. As one can see, except
for the $N=5$ case, the $\langle m_1\rangle_s/N$ curves
corresponding to different $N$'s merge and become one universal
function of $s/N$. In the beginning this function decays from 1
exponentially like that in the $M=2$ case. However, this curve
does not decay to its stationary value $1/M$ directly, instead,
before it growing up again during a period it becomes too tiny to
be visible. In practice this tiny value can be ignored thus
hereafter we treat this period as the ``empty-urn" period. The
values of $\langle m_1\rangle_s/N$ in this period for our chosen
cases are of the order $10^{-4}$. The precise duration of the
empty-urn period depends on how accurate we treat the urn as
empty. After this period the curve grows up again and oscillates
several times with evanescent amplitudes and finally approaches
the stationary mean value $1/M$.

Further exploration shows that there can have more than one
empty-urn period. And the total number of these periods, their
durations, and the local maxima and minima of each $\langle
m_1\rangle_s/N$ curve are all determined by $M$. Fig.~\ref{mean2}
shows the results for $N=60$ at $M=30,60,90$. These three $M$'s
are chosen intentionally to have a simple integer proportion
$1:2:3$ in order to help us inspect relevant results without much
effort. The first maxima of $\langle m_1\rangle_s/N$ for these
three $M$'s are located at $s/N=30,60,90$, which are exactly these
$M$'s. In addition, the locations of the first maxima for $M=60$
and $90$ are just the same locations of the second and third
maxima for $M=30$, respectively. Moreover, for the $M=60$ and
$M=90$ cases the total number of empty-urn periods are more than
one.

Starting from Eq.~(\ref{exp}), we now derive two useful
approximations of $\langle m_1\rangle_s/N$ to help us understand
these observations. First, define \beq \tau\equiv \frac{s}{N}\eeq
and expand the exponential functions in Eq.~(\ref{exp}) as power
series of $\tau$, we have
\bqn
 \hspace{-0.5cm}\frac{\langle m_1\rangle _s}{N}
 &=&
 \frac{e^{-\tau}}{M}\sum^{M}_{j=1}\left(1+\tau q^{-1}_j
 +\frac{1}{2}\tau^2q^{-2}_j+\cdots\right)\non\\
 &=&
 e^{-\tau}\left[1+\frac{\tau^M}{M!}+\frac{\tau^{2M}}{(2M)!}+\cdots\right].
\label{mfactor} \eqn In deriving Eq.~(\ref{mfactor}), we have used
the fact that $\sum^M_{j=1}q^{-k}_j=\sum^M_{j=1}q^{-jk}_1=0$
except for $k=rM$ with $r$ an integer. Note that the form of the
$j$th term ($j>0$) appearing in the last line of
Eq.~(\ref{mfactor}) is the same as the probability of $n$
successful trials in a {\it Poisson process} \cite{Mathew}: \beq
 P(n)=\frac{e^{-\tau}\tau^n}{n!}.\label{poisson}
\eeq Here $n=jM$, and $\tau=\langle n\rangle$ is the expectation
value of $n$. This is not an accident and can be easily
understood. The quantity $\langle m_1\rangle_s/N$ represents not
only the average number of balls in the first urn divided by $N$
but also the probability of finding a certain ball, say, ball 1,
in the first urn. At each time step ball 1 has the probability of
$p=1/N$ being picked out and moved to the next urn. Now since $N$
is large, $p$ is small. In this limit if we do the same operation
$s$ times (here we assume $s$ is also large and define
$\tau=ps=s/N$), then the probability for ball 1 to move $n$ steps
forward ($n$ successful trials) from the first urn is given by
Eq.~(\ref{poisson}). Furthermore, since our system has a
circulating property, the probability of finding ball 1 in the
first urn after $s$ steps consists of the following possibilities:
(1) ball 1 has never been selected (the corresponding probability
is $(1-p)^s\approx e^{-ps}=e^{-\tau}$), and (2) it has been picked
up $M$ times, and (3) it has been chosen $2M$ times, and so on.
The summation of all these contributions gives us the expression
of the last line of Eq.~(\ref{mfactor}).

Applying the {\it saddle-point method} to each
$e^{-\tau}\tau^{jM}$ term and {\it Stirling formula} \cite{Mathew}
to each $(jM)!$ term, we obtain \beq
 e^{-\tau}\tau^{jM}\approx
 \left(\frac{jM}{e}\right)^{jM}e^{-(\tau-jM)^2/2jM},
\eeq
\beq
 (jM)!=\sqrt{2\pi jM}\left(\frac{jM}{e}\right)^{jM}.
\eeq
We thus have
\beq
 \frac{\langle m_1\rangle_s}{N}\approx
 e^{-\tau}+\frac{e^{-(\tau-M)^2/2M}}{\sqrt{2\pi
 M}}+\frac{e^{-(\tau-2M)^2/4M}}{\sqrt{4\pi
 M}}+\cdots.\label{sumgauss}
\eeq

Eq.~(\ref{sumgauss}) tells us that: (1) The $\langle
m_1\rangle_s/N$ curve consists of an exponentially decaying term
$e^{-\tau}$ and a series of Gaussian terms of different heights.
(2) The center of the $j$th Gaussian term is located at $jM$,
which is also a good approximation to the location of the $j$th
local maximum of the $\langle m_1\rangle_s/N$ curve. The
separation between two successive maxima is thus $\Delta \tau=M$.
(3) The height and the standard deviation of the $j$th Gaussian
term are $1/\sqrt{2\pi jM}$ and $\sqrt{jM}$, respectively. (4) If
$j<<M$ we have $\sqrt{jM}<<\Delta \tau$, and the overlap between
two successive Gaussian terms can be neglected. In this situation
the $\langle m_1\rangle_s/N$ curve in the neighborhood of $jM$ can
be well approximated by the $j$th Gaussion term. Similarly, the
$j$th local minimum between the $j$th and $(j+1)$th local maxima
are approximated by \beq \left( \frac{1}{\sqrt{
 j}}+\frac{1}{\sqrt{(j+1)}}\right)\frac{e^{-M/4(2j+1)}}{\sqrt{2\pi M}}.
\eeq

When $j$ is too large, Eq.~(\ref{sumgauss}) becomes inaccurate,
and we must find another expression for better descriptions. Note
that Eq.~(\ref{exp}) can be rewritten as \bqn \frac{\langle
m_1\rangle_s}{N}&=&\frac{1}{M}
\left[1+2e^{-\tau(1-\cos\theta_1)}\cos(\tau\sin\theta_1)\right.\non\\
&&+\left.2
e^{-\tau(1-\cos\theta_2)}\cos(\tau\sin\theta_2)+\cdots\right],\eqn
where we have defined \beq \theta_j=\frac{2\pi
j}{M}=j\theta,\;\;\;\;\theta=\frac{2\pi}{M}.\eeq

Remember that $M$ is a large number, so
\beq
 \sin\theta_1\approx
 \theta,\;\;\;1-\cos\theta_1\approx\frac{\theta^2}{2}.
\eeq
We thus have the approximation
\beq
 \frac{\langle m_1\rangle_s}{N} \approx\frac{1+2
 e^{-\tau\theta^2/2}\cos\left(\tau \theta\right)+
 2e^{-2\tau\theta^2}\cos\left(2\tau
 \theta\right)}{M}\label{alternative}
\eeq for describing the behaviors of $\langle m_1\rangle_s/N$
curve in the region $\tau>>M^2/2\pi^2$, that is,
$\tau(1-\cos\theta_1)\approx \tau\theta^2/2>>1$. From this
expression we also get the result \beq
 \lim_{s\rightarrow \infty}\langle m_1\rangle _s = \frac{N}{M},
\eeq which must be true after the system has reached its
stationary state.

\begin{figure}[hbt]
\epsfxsize=2.8in\epsffile{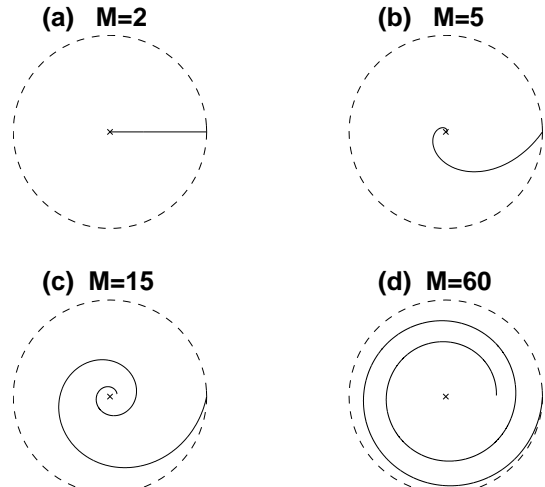} \caption{Plot of
the trace of the center of mass (COM) as a function of time step
$s$, assuming initially all the balls are in the first urn. Here
the curves are plotted for $s=0$ to $s=2MN$ with $M=2,5, 15, 60$.
Each curve with large enough $M$ ($M\geq 10$) has circulated the
origin (the ``x" symbol) of the complex plane twice after evolving
$2MN$ steps.} \label{spiral}
\end{figure}

Up to now we have been focusing our attention on only one single
urn. In fact, we can also understand the behaviors of the system
in a global manner. First, let's define the ``phase angle" of the
$k$th urn (see Fig.~\ref{arg}) as  \beq
 \phi_k=-(k-1)\,\theta.\label{phik}
\eeq Now we define the ``center-of-mass" (COM) of the $N$-ball,
$M$-urn system as \beq
 \mbox{COM}\equiv\sum^M_{k=1} e^{i\phi_k} \langle
 m_k\rangle_s.\label{com}
\eeq

\begin{figure}[hbt]
\epsfxsize=2.8in\epsffile{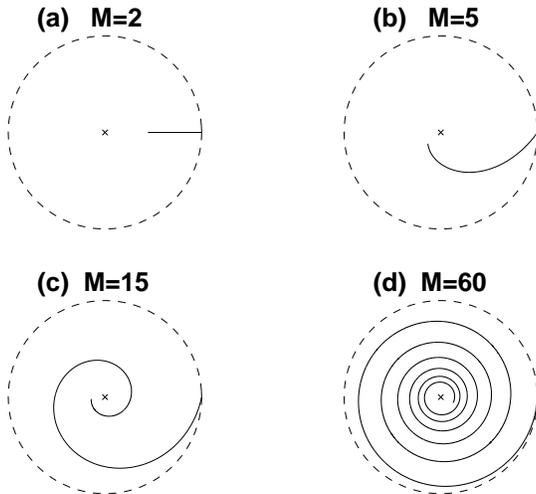} \caption{These COM
curves are plotted for $s=0$ to $s=M^2N/{\pi}^2$ with $M=2,5, 15,
60$. The norm of a $\mbox{COM}$ in the complex plane becomes
$e^{-2}=0.1353$ times smaller after evolving $M^2N/{\pi}^2$ steps
if $M$ is large enough ($M\geq 10$). Here the symbol ``x" denotes
the origin.} \label{spiral1}
\end{figure}

According to Eq.~(\ref{com}) the COM is in general a complex
number, say, $Z=Re^{i\phi}$. Here $R=|Z|$ satisfies $0\leq R\leq
N$, and $\phi$ is the phase angle of $Z$. The variation of the the
phase angle $\phi$ with respect to $s$ or $\tau=s/N$ represents
how fast on average these $N$ balls circulate around the $M$ urns,
and the norm $R$ here gives us the information of the extent of
the distribution of the $N$ balls. If we consider the normalized
center-of-mass ${\rm COM}/N\equiv re^{i\phi}$ instead of COM, we
conclude that the trace of $re^{i\phi}$ must be contained inside
the unit circle. Substitute (\ref{mmean}) and (\ref{phik}) into
(\ref{com}) and use (\ref{lambdaq}) and (\ref{qm}), we get \bqn
 \frac{\mbox{COM}}{N}&=&\frac{1}{M}\sum^M_{j=1}\sum^M_{k=1}q^{(j-1)(k-1)}_1
 \lambda^s_j\non\\&=&\lambda^s_1\non\\
 &\approx&
 e^{-\tau\theta^2/2}e^{-i\tau\theta},\label{com1}
\eqn hence
\beq
 r=e^{-\tau\theta^2/2},\;\;\;\;\phi=-\tau\theta.\label{rphi}
\eeq Here we see that the ${\rm COM}/N$ curve is approximately
described by a spiral circulating inside a unit circle. The
angular frequency of the circulating motion of the COM with
respect to $s$ is $-\theta=-2\pi/M$ (clockwise), consistent with
both the definition of our game and the oscillation behaviors of
the ${\rm COM}/N$ curve discussed before. Furthermore, when
$\tau>M^2/2\pi^2$ we have $r<e^{-1}$, which indicates that the
system has a wide extent of the ball-distribution, also consistent
with the previous discussions. Some examples are illustrated in
Fig~\ref{spiral} and Fig.~\ref{spiral1}.

\section{State matrix and generating function}

Now we calculate $\langle {\bf m}|P^s|{\bf m}_0\rangle$ -- the
transition probability from $|{\bf m}_0\rangle$ to $|{\bf
m}\rangle$ after $s$ steps. Once one knows the exact solution of
$\langle {\bf m}|P^s|{\bf m}_0\rangle$, any quantity can be
calculated explicitly.

Define \beq S_{{\bf m}{\bf m}'}=\langle {\bf m}|P|{\bf m}'\rangle,
\eeq then we have: \beq
    \langle {\bf  m}|P^s|{\bf  m}_0\rangle  =    (S^s)_{{\bf  m}{\bf  m}_0}.
\eeq Here $S$ is a $ H_N^M \times H_N^M $ matrix, we name it {\it
state matrix}, and $|{\bf  m}\rangle $ is a $H_N^M$ column vector,
here \beq
     H_N^M = C_{M-1}^{N+M-1} = \frac{(N+M-1)!}{N!(M-1)!}.
\eeq Like before, $S^s$ can be calculated by means of its
eigenvalues and eigenstates. According to Eq.~(\ref{recu}), the
matrix $S$ has components \bqn
             S_{{\bf  m}{\bf  m}'}&=& \sum_{i=1}^{M}\frac{m_i+1}{N}
                      \delta_{m_1,m'_1}\delta_{m_2,m'_2}\cdots  \non\\
                      &&\times\delta_{m_i+1,m'_i}\delta_{m_{i+1}-1,m'_{i+1}}
                      \cdots \delta_{m_M,m'_M},
\eqn where $m_{M+1}=m_1$ have been assumed. The eigenvalue
equation can be written as \beq
   \sum_{\{{\bf  m}'\}}S_{{\bf  m}{\bf  m}'} \phi_{{\bf  m}'}=\gamma
   \phi_{{\bf  m}},\label{sphi}
\eeq or more explicitly \bqn
  \sum_{i=1}^{M}&&\hspace{-5mm}\frac{m_i+1}{N}\phi_{m_1,m_2,\cdots,m_i+1,m_{i+1}-1,\cdots,m_M}\non\\
  &=&\gamma \phi_{m_1,m_2,\cdots,m_i ,m_{i+1}
  ,\cdots,m_M}.\label{Seq}
\eqn In the above expression we have set
$\phi_{m_1,m_2,\cdots,m_M}=0$ for those $|{\bf
m}\rangle=|m_1,m_2,\cdots,m_M\rangle$ that do not satisfy the
constraint (\ref{constraint}).

It is not an easy task to diagonalize $S$ directly. Thus we adopt
another strategy. We first construct a generating function for
$\phi_{m_1,m_2,\cdots,m_M}$ and then transform the matrix
eigenvalue equation (\ref{sphi}) to its differential equation
form. We find that the differential equation can be solved
analytically.

By introducing variables $x_1,x_2,\cdots,x_M$, the generating
function can be defined as
\beq
  f(x_1,x_2,\cdots,x_M)\\
  \equiv \sum_{\{{\bf  m}\}}  \phi_{m_1,m_2,\cdots,m_M}
                  x_1^{m_1}x_2^{m_2}\cdots x_M^{m_M}.
\eeq Hereafter we also use the following expression \beq
  f({\bf X})= \sum_{\{{\bf  m}\}}  \phi_{{\bf  m}}  X^{{\bf
  m}},\label{finshort}
\eeq where ${\bf X}$ and $X^{\bf m}$ are defined by
\beq
 {\bf X }\equiv\left[\bary{ccccc}x_1\\x_2\\
 \vdots\\x_M\eary\right],\;\;\; X^{{\bf  m}}\equiv
 x_1^{m_1}x_2^{m_2}\cdots x_M^{m_M}.
\eeq

To proceed further, note that $f({\bf X})$ satisfies the following
two relations:
\bqn
 \partial_{x_i}f({\bf X})&=&\sum_{\{\bf m\}} (m_i+1)
 \phi_{m_1,m_2,\cdots,m_i+1, \cdots,m_M}  X^{{\bf
 m}},\non\\
 x_if({\bf X})&=&\sum_{\{\bf m\}}  \phi_{m_1,m_2,\cdots,m_i-1, \cdots,m_M}
 X^{{\bf  m}},\label{twof}
\eqn  as can be easily checked. Multiplying $X^{\bf m}$ on both
sides of Eq.(\ref{Seq}), summing over all $\{\bf m\}$, and using
the results of (\ref{twof}) , we get
\beq
  \sum_{i=1}^{M}\frac{x_{i+1}}{N} \partial_{x_i}f({\bf X})=
  \gamma f({\bf X}),
\eeq or equivalently \beq
  \sum_{i=1}^{M}x_{i+1}\partial_{x_i} \ln[f({\bf X})]= N
  \gamma,\label{lnf}
\eeq which is the desired differential equation form of the
eigenvalue equation (\ref{Seq}).
Define
\beq
 x_{q_j^{} }=x_1q_j^{} +x_2q_j^2+ \cdots +x_Mq_j^M,
\eeq
we find
\beq
  \sum_{i=1}^{M}x_{i+1}\partial_{x_i} \ln(x_{q_j})=q^{-1}_j=q^*_j.\label{lnxq}
\eeq This implies that the complete solution of $\ln[f({\bf X})]$
can be written as
\beq
  \ln[f_{{\bf  n}}({\bf X})] = \sum_{j=1}^{M} n_j \ln(x_{q_j^{} }),
\eeq which gives us
\beq
 f_{{\bf  n}}({\bf X}) =
 \prod_{j=1}^{M}x_{q_j^{} }^{n_j}\equiv X^{\bf n}_{\bf q}.
\eeq Here $f_{{\bf  n}}({\bf X})$(a homogeneous $N$th power
function) and the eigenvalue $\gamma_{\bf n }$ are characterized
by ${\bf n}=[n_1,n_2,\cdots,n_M]$ and $ {\bf
q}=[q_1,\,q_2,\cdots,\,q_M]$, satisfying \beq
 N = \sum_{j=1}^{M} n_j,\;\;\;\;
  \gamma_{{\bf  n}} = \frac{1}{N}\sum_{j=1}^{M} n_jq_j^{\ast}
  = \frac{{\bf  n}\cdot {\bf  q}^{\ast}}{N}.\label{nandga}
\eeq

Denoting the ${\bf n}$-th eigenvector of $S$ as $\phi({\bf n})$,
Eq.~(\ref{sphi}) and (\ref{finshort}) now become \beq
 \sum_{\{{\bf  m}' \}}S_{{\bf  m}{\bf  m}' }
 \phi_{{\bf  m}'}({\bf  n}) = \gamma_{{\bf  n}}
 \phi_{{\bf  m}}({\bf  n})
\eeq and \beq
  f_{\bf n}({\bf X})= \sum_{\{{\bf  m}\}}  \phi_{{\bf  m}}({\bf n})  X^{{\bf
  m}}=X^{\bf n}_{\bf q}.\label{finshort1}
\eeq

To diagonalize $S$ we first define an orthogonal transformation
matrix $U$: \beq
     U_{{\bf  m}{\bf  n}} =  \phi_{{\bf  m}}({\bf  n}),
\eeq where $ \phi_{{\bf  m}}({\bf  n})$ according to
Eq.~(\ref{finshort1}) is the coefficient of $ X^{{\bf  m}}$ that
appears in the expansion of $f_{{\bf n}}({\bf X})=X^{\bf n}_{\bf
q}$.

We now are ready to solve the matrix $U^{-1}$. Multiplying $X^{\bf
m}=x_1^{m_1}x_2^{m_2} \cdots x_M^{m_M}$ on both sides of \beq
 \sum_{\{{\bf  n}\}}
 \phi_{{\bf  m}}({\bf  n})\, U_{{\bf  n}{\bf  l}}^{-1}
 =\sum_{\{{\bf  n}\}} U_{{\bf  m}{\bf  n}}\,U_{{\bf  n}{\bf  l}}^{-1}=
 \delta_{{\bf  m}{\bf  l}}
\eeq and summing over all possible ${\{{\bf m}\}}$, we get \beq
 \sum_{\{{\bf  n}\}} f_{{\bf  n}}({\bf X}) \,U_{{\bf  n}{\bf m}}^{-1}
= \sum_{\{{\bf  n}\}} X^{\bf n}_{\bf  q}\, U_{{\bf  n}{\bf
m}}^{-1}= X^{\bf  m}. \label{fux}\eeq

Furthermore, define two vectors ${\bf Y}$ and ${\bf X}_{\bf q}$ as
\bqn
 {\bf Y}&=&
 \left[ \bary{cccc} y_1 & \\ y_2 & \\ \vdots &
 \\ y_M & \eary \right]
 \equiv \left[ \bary{cccc} q_1    &  q_2   & \cdots &
 q_m \\
 q_1^2  &  q_2^2    &  \cdots & q_m^2 \\
 \vdots &  \vdots   & \ddots & \vdots \\
 q_1^M  &  q_2^M    & \cdots & q_m^M
 \eary \right] \left[ \bary{cccc} x_1 & \\ x_2 & \\
 \vdots & \\ x_M & \eary \right],
\eqn and \beq {\bf X}_{\bf q}\equiv
 \left[ \bary{cccc} x_{q_1} & \\ x_{q_2} & \\  \vdots
 & \\ x_{q_M} & \eary \right],
\eeq we have \beq {\bf Y}=\sqrt{M}R\, {\bf X}.\eeq

Using the same notation and remember that $R^{-1}=R^{\ast}$, we
find \beq
  {\bf X} = \frac{1}{\sqrt{M}}
 R^{\ast} {\bf Y} = \frac{1}{M}{\bf Y}_{{\bf q}^{\ast}}.
\eeq These results further lead to
\bqn
 X^{{\bf  m}} & = & \frac{1}{M^N}
( Y_{{\bf q}^{\ast}})^{{\bf  m}}
      = \frac{1}{M^N} Y_{\bf q}^{\tilde{{\bf  m}}}
      = \frac{1}{M^N}f_{\tilde{{\bf  m}}}({\bf Y})  , \non \\
     &=&  \frac{1}{M^N} \sum_{\{{\bf n}\}} \phi_{{\bf  n}}\left(\,\tilde{{\bf  m}}\,\right)
      Y^{{\bf n}}= \frac{1}{M^N} \sum_{\{{\bf n}\}} \phi_{{\bf  n}}\left(\,\tilde{{\bf  m}}\,\right)
      X^{{\bf n}}_{\bf q} \non\\
      &=&  \frac{1}{M^N} \sum_{\{{\bf n}\}} f_{{\bf  n}}({\bf X}) \phi_{{\bf  n}}
      \left(\,\tilde{{\bf  m}}\,\right).\label{many}
\eqn Comparing Eq. (\ref{many}) with (\ref{fux}), we get: \beq
  U_{{\bf  n}{\bf  m}}^{-1} = \frac{1}{M^N}    \phi_{{\bf  n}}\left(\,\tilde{{\bf  m}}\,\right),
\eeq where we have used the relations:
\bqn
     Y_{{\bf q}^{\ast}}^{{\bf  m}}
 & = & y_{q_1^{\ast}}^{m_1} y_{q_2^{\ast}}^{m_2}y_{q_3^{\ast}}^{m_3} \cdots
       y_{q_{M-1}^{\ast}}^{m_{M-1}}y_{q_M^{\ast}}^{m_M} , \non \\
 & = & y_{q_{M-1}}^{m_1} y_{q_{M-2}}^{m_2}y_{q_{M-3}}^{m_3} \cdots
         y_{q_1}^{m_{M-1}}y_{q_M}^{m_M} , \non \\
 & = & y_{q_1}^{m_{M-1}} y_{q_2}^{m_{M-2}}y_{q_3}^{m_{M-3}} \cdots
         y_{q_{M-1}}^{m_1}y_{q_M}^{m_M} , \non \\
 & = &  Y^{\tilde{\bf m}}_{\bf q}= f_{\tilde{{\bf  m}}}({\bf Y}),
\eqn and defined $\tilde{\bf  m}$ as \beq
     \tilde{\bf  m} \equiv [ m_{M-1},m_{M-2}, m_{M-3}, \cdots,  m_1, m_M
     ].\label{tilde}
\eeq

Finally we obtain the desired solution of $\langle {\bf
m}|P^s|{\bf m}_0\rangle $:
\bqn
    \langle {\bf  m}|P^s|{\bf  m}_0\rangle  &=&  (S^s)_{{\bf  m}{\bf  m}_0},   \non  \\
              &=& (U \Gamma^s U^{-1})_{{\bf  m}{\bf  m}_0},  \non  \\
              &=&  \frac{1}{M^N} \sum_{{\bf  m}'} \gamma_{{\bf  m}'}^s
  \phi_{{\bf  m}}({\bf  m}') \phi_{{\bf  m}'}(\tilde{{\bf  m}}_0),
\eqn where $\Gamma$ is the eigenvalue matrix of $S$, which has
components $\Gamma_{{\bf m}{\bf m}'}=\gamma_{{\bf m}} \delta_{{\bf
m}{\bf m}'}$.

\section{poincar\'{e} cycle}
In this section, we study the Poincar\'{e} cycle of our periodic
urn model. For simplicity we first consider the situation that
initially all the $N$ balls are stayed in the last urn, i.e.,
\beq
  {\bf  m}_0 = [0,0,0, \cdots ,N].\label{m0} \non
\eeq From (\ref{m0}) and (\ref{tilde}) we have
\beq
 {\bf m}_0=\tilde{\bf m}_0.
\eeq

Now we want to know how many time steps on average are required
for all of the $N$ balls to return to the last urn (the initial
state). We thus have to calculate
\beq
    \langle {\bf  m}_0|P^s|{\bf  m}_0\rangle
              =  \frac{1}{M^N} \sum_{{\bf  m}} \gamma_{{\bf  m}}^s
  \phi_{{\bf  m}_0}({\bf  m}) \phi_{{\bf  m}}(\tilde{{\bf  m}}_0).
\eeq

Recall that $\phi_{\bf m}({\bf n})$ is nothing but the coefficient
of $X^{\bf m}=x^{m_1}_1\cdots x^{m_M}_M$ appearing in the
expansion of $f_{\bf n}({\bf X})=X^{\bf n}_{\bf
q}=x^{n_1}_{q_1}\cdots x^{n_M}_{q_M}$. From Eq.~(\ref{finshort1})
and (\ref{m0}), we have \beq
  \phi_{{\bf  m}_0}({\bf  m})  = 1,
\eeq
\beq
 \phi_{{\bf  m}}(\tilde{{\bf  m}}_0)  =  \frac{N!}{m_1!m_2!m_3! \cdots m_M!} \equiv
 \left( \bary{cc}  N  \\ {\bf  m}    \eary \right),
\eeq and hence \beq
 \langle {\bf  m}_0|P^s|{\bf  m}_0\rangle =
 \frac{1}{M^N} \sum_{{\bf  m}} \left( \bary{cc} N  \\ {\bf  m}
 \eary \right) \gamma_{{\bf  m}}^s \equiv {\cal P}(s).\label{ps}
\eeq Here ${\cal P}(s)$ represents the transition probability for
the system to return to the initial state after $s$ steps. It does
not preclude the possibility that the initial state has already
been re-arrived before.

Since Poincar\'{e} cycle is defined as the time interval required
for the event of first return to happen, so we have to do more
calculations to extract what we really want. We define a function
${\cal Q}(s)$ as the probability for the event of first return to
happen at the $s$th step. The Poincar\'{e} cycle can thus be
defined as \beq {\rm P}.\,{\rm C}.=\sum^{\infty}_{s=0}s{\cal
Q}(s). \eeq

By definition ${\cal Q}(s)$ relates to ${\cal P}(s)$ via the
relation \beq
 {\cal P}(s) = {\cal Q}(s) +
 \sum_{k=1}^{s-1}{\cal Q}(k){\cal P}(s-k),\label{qps}
\eeq and hence ${\cal Q}(s)$ can be calculated from ${\cal P}(s)$.
To ease the calculation we now use again the generating function
method. We first define two generating functions:

\beq
  h(z) \equiv  \sum_{s=1}^{\infty}{\cal P}(s)z^s,\;\;\;\;
  g(z) \equiv  \sum_{s=1}^{\infty}{\cal Q}(s)z^s,\label{hz}
\eeq and then we find from Eq.~(\ref{qps}) that \beq
 g(z) = \frac{h(z)}{h(z)+1}.
\eeq These two generating functions also lead to
\beq
     \sum_{s=0}^{\infty} s {\cal Q}(s) =
     \left( \frac{dg}{dz} \right)_{z=1},\;\;\;\;
 g' =  \frac{h'}{(1+h)^2},
\eeq which can determine the Poincar\'{e} cycle.

We now calculate $h(z)$. From Eq.~(\ref{ps}) and (\ref{hz}) we
obtain \bqn
  h(z) & = & \frac{1}{M^N} \sum_{\bf  m}
             \left( \bary{cc} N \\ {\bf  m}\eary \right)
             \sum_{s=1}^{\infty} \left( \gamma_{\bf  m} z \right)^s, \non \\
       & = & \frac{1}{M^N} \sum_{{\bf  m}}
             \left( \bary{cc} N  \\ {\bf  m}    \eary \right)
              \left( \frac{ \gamma_{{\bf  m}}z }{1-\gamma_{{\bf  m}}z} \right).
\eqn Since we know from Eq.~(\ref{nandga}) that $\gamma_{{\bf
m}_0}=1$, thus when $z \rightarrow 1^-$ , $h(z)$ becomes singular:
\beq
 \lim_{z \rightarrow 1^-}h(z) = \frac{1}{M^N} \frac{z}{1-z} +
 \mbox{regular function}.
\eeq In this limit, we obtain \beq \lim_{z \rightarrow 1^-}g'=
\lim_{z \rightarrow 1^-}\frac{h'}{(1+h)^2} = M^N, \eeq which gives
us the desired Poincar\'{e} cycle: \beq {\rm P}.\,{\rm
C}.=\sum_{s=0}^{\infty} s {\cal Q}(s) = M^N. \eeq

To understand this result, we refer to the {\it ergodic theorem}
\cite{Huang}, which says that if one waits a sufficiently long
time, the locus of the representative point of a system will cover
the entire accessible phase space. For our periodic urn model, the
``representative point" corresponds to the microstate of the
arrangement of balls, the ``accessible phase space" is the set of
total microstates, and the ``locus" means the evolution history of
the system (See Fig.~\ref{arg}). Since in our periodic model every
microstate has the same probability to appear (a fundamental
assumption of statistical mechanics), and the set of total
microstates contains $M^N$ microstates, therefore on average one
has to wait $M^N$ time steps to see a microstate reappearing
again.

What will be the Poincar\'{e} cycle if the initial state is
different from the ${\bf m}_0$ in Eq.~(\ref{m0})? Let's denote the
initial state by ${\bf d}$:
\beq
 |{\bf d}\rangle=|d_1,d_2,\ldots,d_M\rangle.
\eeq Now we have \beq {\cal P}(s)= \langle {\bf d}|
 P^s|{\bf d}\rangle=\frac{1}{M^N}\sum_{\{\bf m\}}\gamma^s_{\bf m}\phi_{\bf
 d}({\bf m})\phi_{\bf m}(\tilde{\bf d}),
\eeq and
\beq
 h(z)=\frac{1}{M^N}\sum_{\{\bf m\}}\phi_{\bf
 d}({\bf m})\phi_{\bf m}(\tilde{\bf d})
 \left(\frac{\lambda_{\bf m}z}{1-\lambda_{\bf m}z}\right).
\eeq

In general it is difficult to calculate $\phi_{\bf
 d}({\bf m})$ and $\phi_{\bf m}(\tilde{\bf d})$. However, we do not need to
 calculate them all. Remember that to determine the Poincar\'{e}
 cycle the the knowledge of the asymptotic form of $h(z)$
 near $z=1$ is enough. This is given by
\beq
 \lim_{z\rightarrow 1^{-}}h(z)=\frac{\phi_{\bf
 d}({\bf m}_0)\phi_{{\bf m}_0}(\tilde{\bf d})}{M^N}
 \frac{z}{1-z}+{\rm regular}\;{\rm function}.\label{hz}
\eeq
Substituting
\beq
 \phi_{\bf d}({\bf m}_0)=\left( \bary{cc} N \\ {\bf
 d}\eary \right),\;\;\;\;\;
 \phi_{{\bf m}_0}(\tilde{\bf d})=1
\eeq into (\ref{hz}), we find \beq
 {\rm P. C.}=\lim_{z\rightarrow 1^{-}}\frac{h'}{(1+h)^2}=\frac{M^N}{\left(\bary{cc} N \\ {\bf d}\eary\right)}
\eeq This result conforms with our intuition. Consider the
difference between the concepts of {\it configuration} and {\it
microstate} , one can easily understand this result because the
degeneracy of the configuration ${\bf d}$ is given by
\[
 \left(\bary{cc} N\\ {\bf d}\eary\right)=\frac{N!}{d_1 !d_2!\cdots d_M !}.
\]

\section{summary}

In this work we propose a generalized Ehrenfest urn model of many
urns arranged periodically along a circle. We solve an $N$-ball,
$M$-urn problem explicitly. The evolution of the system is
studied, and the average number of balls in a certain urn at any
time step is calculated. We find that this mean value oscillates
several times before it arrives the stationary value. We also
obtained the Poincar\'{e} cycle. The result is consistent with the
consideration of the total number of possible microstates of the
system.

\section*{Acknowledgments}
The authors would like to acknowledge helpful discussions with
professor T. F. Jiang. This work received support from the
National Science Council, Republic of China through Grant No. NSC
90-2811-M-009-019.


\begin{references}

\bibitem{Huang}
K. Huang, {\it Statistical Mechanics} (John Wiley \& Sons, Inc,
1987).

\bibitem{Erf}
P. Ehrenfest and T. Ehrenfest, Physik. Z. {\bf 8}, 311 (1907).

\bibitem{Siegert}
A. J. F. Siegert, Phys. Rev. {\bf 76}, 1708 (1949).

\bibitem{Klein}
M. J. Klein, Phys. Rev. {\bf 103}, 17 (1956).

\bibitem{Calvo}
J. G\"{u}\'{e}mez, S. Velasco, and A. Calvo Hern\'{a}ndez, Am. J.
Phys. {\bf 57} , 828 (1989).

\bibitem{God}
C. Godr\`{e}che, J. P. Bouchaud, and M. M\'{e}zard, J. Phys. A:
Math. Gen. {\bf 28}, L603 (1995).

\bibitem{Kanter}
R. Metzler, W. Kinzel, and Kanter, J. Phys. A: Math. Gen. {\bf
34}, 317 (2001).

\bibitem{Lipo}
A. Lipowski, J. Phys. A: Math. Gen. {\bf 30}, L91 (1997).

\bibitem{Arora}
D. Arora, D. P. Bhatia, and M. A. Prasad, Phys. Rev. E {\bf 60} ,
145 (1999).

\bibitem{Ritort}
F. Ritort, Phys. Rev. Lett. {\bf 75} , 1190 (1995).

\bibitem{Eggers}
J. Eggers, Phys. Rev. Lett. {\bf 83} , 5322 (1999).


\bibitem{Na}
Devaraj van der Meer, Ko van der Weele, and Detlef Lohse, Phys.
Rev. E {\bf 63} , 061304 (2001).

\bibitem{van}
Devaraj van der Meer, Ko van der Weele, and Detlef Lohse, Phys.
Rev. Lett. {\bf 88} , 174302 (2002).

\bibitem{Droz}
A. Lipowski and M. Droz, Phys. Rev. E {\bf 65} , 031307 (2002).

\bibitem{Mathew}
Jon Mathews,  Robert L. Walker, {\it Mathematical Methods of
Physics}, 2nd. ed. (Addison Wesley, 1971)


\end{references}
\end{document}